\documentclass[epj]{svjour}
\usepackage{graphicx}
\usepackage{graphicx}
\usepackage[usenames,dvipsnames]{color}
\usepackage{amsmath,amssymb}
\usepackage{epstopdf}
\usepackage{hyperref}
\begin{document}
\title{Generation of Magnetic Fields Near QCD Transition by Collapsing $Z(3)$
  Domains}
\author{Abhishek Atreya\inst{1} \and Soma Sanyal\inst{2}
}                     
%
%
\institute{Center For Astroparticle Physics and Space Sciences, Bose
  Institute, Kolkata, 700091, India. \email{atreya.abhi@gmail.com} \and
  School of Physics, University of Hyderabad Telangana, 500046, India.
  \email{sossp@uohyd.ernet.in}}
\date{Received: / Revised version:}
%
\abstract{
    In this work we investigate the possibility of generation of primordial
  magnetic field in the early universe near the QCD phase transition epoch via
  the collapse of $Z(3)$ domains. The $Z(3)$ domain walls arise in the
  deconfined phase of the QCD (above $T\sim 200$ MeV) and their collapse leads
  to a net quark concentration near the wall boundary due to non-trivial
  reflection of quarks. We look at the response of leptons to this quark
  excess and find that leptons do not cancel the electric charge concentration
  due to the quarks. The wall collapse and a net charge concentration can lead
  to the generation of vorticity and turbulence in the primordial plasma. We
  estimate the magnitude of the magnetic field generated and find that it can
  be quite large $\mathcal{O}(10^{15-18}\mathrm{G})$. The mechanism is
  independent of the order of the QCD phase transition.
} 
\maketitle

\section{INTRODUCTION}
\label{sec:intro} 

The origin of the observed magnetic fields in the Universe remains an
unresolved problem of cosmology. Observations
like Zeeman splitting of spectral lines, polarization and
  intensity measurement of of synchrotron radiation of electrons and Faraday
  rotation measurements indicate the presence of
large scales magnetic fields in the universe. The strengths of these fields
vary from $1-10~\mu$G for the length scales of the order of a few kpc to
strengths of $\mathcal{O}(10^{-15}-10^{-18}\mathrm{G})$ for Mpc scales.
The kpc scale field correlations can be explained by producing the seed 
magnetic fields using a Biermann battery mechanism in the proto galaxy, which 
are then amplified by a galactic dynamo. This process, however, fails for 
the large scale magnetic fields. An appealing alternative to the above 
scenario is to argue that the seed magnetic field has primordial origin which 
gets amplified as the proto galactic cloud undergoes collapse. This would 
naturally provide magnetic fields at all scales. For details,
  we refer to ref.
  \cite{Grasso:2000wj,Widrow:2002ud,Widrow:2011hs,Subramanian:2015lua}
  and the references cited therein.

The Universe has a rich thermal history and each of the stage can provide us
the seed required to produce the observed magnetic field. In this work we
focus on the epoch of the Quark-Hadron (Q-H) transition. This is expected to
occur when the universe was roughly micro-seconds old. Till the turn of the
century the Q-H transition was thought to be a first order transition. The
bubble wall dynamics associated 
with the first order transition provided a rich spectrum of possibilities 
like quark nuggets as dark matter candidates \cite{Witten:1984rs}, 
production of baryon inhomogeneities \cite{Fuller:1987ue} and also the 
primordial magnetic fields \cite{Quashnock:1988vs,Cheng:1994yr,Sigl:1996dm}.
Unfortunately, none of the above scenarios hold in the light of results from 
lattice gauge theory showing that a first order quark-hadron transition
is very unlikely. The transition, for the range of chemical potentials
relevant for the early universe, is most likely a crossover. Thus there are
no bubble walls and hence the entire spectrum of possibilities is
relinquished. See ref. \cite{DeTar:2009ef,Andersen:2014xxa,DElia:2018fjp} and
references therein for details on the discussion of the order of the phase
transition.

Our proposal for an alternate mechanism for the magnetic field generation is 
through the collapse of closed domain walls in the quark-gluon plasma (QGP)
phase of QCD. An important point to note is that this mechanism does not
depend on the order of the QCD phase transition as the domain walls are
present in the QGP phase of the system and are not a result of the Q-H
transition. Since these defects are in the deconfined phase of QCD and vanish
in the confined phase, they are not constrained, unlike the GUT defects.

The possibility of extended topological objects in the
QGP has been extensively discussed in the literature 
\cite{Bhattacharya:1992qb,West:1996ej,Boorstein:1994rc}. These are domain
wall defects that arise from the spontaneous breaking of $Z(3)$ symmetry 
in the high temperature phase (QGP phase) of QCD.
We should mention that the existence of these $Z(3)$ walls becomes a
non-trivial issue in the presence of quarks \cite{Smilga:1993vb,Belyaev:1991np}.
However it has also been argued that the effect of quarks can be understood
in terms of the explicit breaking of $Z(3)$
symmetry \cite{DeGrand:1983fk,Dumitru:2003cf}. This
interpretation allows for very interesting possibilities for the QGP phase
with rich phenomenology which can have very important implications for
cosmology. This finds support in the recent lattice calculations of QCD with
quarks \cite{Deka:2010bc} and also in the effective model studies of QGP
\cite{Mishra:2016ipq} which suggest that there
is a strong possibility of existence of these $Z(3)$ vacua at high temperature.

However the magnetic field produced during processess that
happen before the QCD phase transition, like the electro-weak phase transition, 
might be present at the time of the Q-H transition (see reviews
\cite{Grasso:2000wj,Widrow:2002ud,Widrow:2011hs,Subramanian:2015lua} and
ref. \cite{Dolgov:2010gy,Ayala:2017gqa} for more recent works).
In presence of magnetic field the QCD phase transition shows intriguing
behaviour. The chiral condensate at $T = 0$ grows monotonically with the
magnetic field strength, an effect known as magnetic catalysis. At finite $T$
the situation gets a bit involved. Below critical temperature, the chiral
condensate increases with magnetic field with values consistently smaller than
those at $T=0$ for the same magnetic field strength. As $T$ approaches $T_{c}$,
the chiral condensate reaches a peak value for a threshold value of magnetic
filed and shows a reduction with further increase in the magnetic field. Above
the critical temperature the chiral condensate decreases for all values of
magnetic field. This phenomenon is termed as the inverse magnetic catalysis
\cite{Andersen:2014xxa}. The effective models of QCD phase transition do not
reproduce this pattern quite well. Even though attempts have been made to
explain the discrepancy between Lattice studies and the effective models by
trying to separate the contributions of valance and sea quarks (see sec.
IXA of ref. \cite{Andersen:2014xxa}), a final word still remains to be said. 
The magnetic field, through quark contribution, tends to break the $Z(3)$
symmetry explicitly \cite{Mizher:2010zb}.
Recent lattice results \cite{Endrodi:2015oba} indicate that in the presence of
an asymptotically large magnetic field, the transition is quite sharp (possibly
a first order too) and is at a lower temperature than in the absence of a
magnetic field. A similar effect is expected in the case of electro-weak phase
transition in presence of magnetic field \cite{Ayala:2004dx,Sanchez:2006tt}.
In such a situation it is quite possible that all the possibilities
mentioned above are realised in the early universe near the Q-H transition. We
refer to \cite{Andersen:2014xxa} and references therein for a comprehensive
discussion of QCD phase transition in a magnetic field.

Even when there is no magnetic field, $Z(3)$ defects can revive the first
order transition scenarios of baryon inhomogeneities and quark nuggets formation
\cite{Layek:2005zu,Atreya:2014sca}.
We might add that these are the only relativistic field theory topological
solitons which are accessible in laboratory experiments. Their detection will
provide deep insights in the non-trivial physics of the QGP phase. It therefore
looks reasonable to explore the possible consequences of these $Z(3)$ domains
and associated walls.

As the $Z(3)$ domain wall collapses, it produces shocks in the plasma which
will lead to turbulence being generated in the wake of the the wall. As the
wall sweeps through the quark gluon plasma, an excess baryon concentration
builds in the collapsing region. We use Poisson's equation to estimate the
electric charge, due to leptons, on the domain wall. We find that the lepton
concentration doesn't cancel the charge due to the baryon concentration.
Assuming the domain wall collapse to be spherical, we calculate the magnetic
field generated during the collapse. The magnetic field generated depends on
the baryon density contrast across the domain wall and the size of the
collapsing region. It can be as high as $10^{15-18}$ G for a baryon density
contrast of the order of $10^6$ within a radius of roughly one meter.
Since the collapse of the domain walls happen before the quark hadron phase
transition, the magnetic field is generated in the quark gluon
plasma epoch. Thus the mechanism is completely independent of the order of
the phase transition. 
 
  The organization of the paper is as follows. In section \ref{sec:z3}
  we briefly discuss the $Z(3)$ symmetry. This is the symmetry of the Polyakov
  loop 
which is the order parameter of the confinement transition. We estimate the
profile of the domain wall and the transmission coefficients for different
quarks.
In section \ref{sec:charge} we discuss the formation of $Z(3)$ structures in
the early universe and how a charge density is accumulated within a collapsing
$Z(3)$ domain In section \ref{sec:magfld}, we first argue that these collapsing
domains can generate the vorticity in the plasma and then make an estimate of
the magnetic field generated. We conclude in section \ref{sec:discussion}.

\section{$Z(3)$ DEFECTS AND QUARK INTERACTIONS}
\label{sec:z3}

\subsection{$Z(3)$ domains in QGP}
 \label{sec:z3qgp} 
 The order parameter of the confinement-deconfinement transition for a pure
 gauge $SU(N)$ system at temperature $T$, is the thermal expectation value
 of the Polyakov loop \cite{Polyakov:1978vu,Gross:1980br,McLerran:1981pb} defined
 as 
\begin{equation} 
  L(x) = \frac{1}{N}\mathrm{Tr}\left\lbrace\mathbf{P}\biggl[\exp\left(ig
      \int_{0}^{\beta}A_{0} (\vec{x},\tau)d\tau\right)\biggr]\right\rbrace,
\label{eq:lx}
\end{equation}
where $\beta = T^{-1}$ and and $g$ is the gauge coupling. The trace denotes
the summing over color degrees of freedom and $\mathbf{P}$ denotes the path
ordering in the Euclidean time $\tau$. The $SU(N)$ gauge fields satisfy the
periodic boundary conditions in $\tau$, viz
$A_{0}(\vec{x},0) = A_{0}(\vec{x},\beta)$.
 
The free energy of a test quark can then be studied by  writing the
partition function and noting that
$\langle L(\vec{x})\rangle \propto e^{-\beta F}$. In the confined phase,
$\langle L(\vec{x})\rangle= 0$ implying that the free energy of a test quark
is infinite (i.e. system is below $T_{c}$). In the deconfined phase, a test
quark has finite free energy and hence $\langle L(\vec{x})\rangle \neq 0$.
We will use $l(x)$ to denote $\langle L(\vec{x})\rangle$ from now on, for the
sake of brevity. Under $Z(N)$ transformation (which is the center of $SU(N)$),
$l(x)$ transforms as
\begin{equation}
l(x) \longrightarrow Z\times l(x), \qquad \text{where}~ Z = e^{i\phi},
\end{equation}
and $\phi = 2\pi m/N$; $m = 0,1 \dotsc (N-1)$. This gives rise to the
spontaneous breaking of $Z(N)$ symmetry with $N$ degenerate vacua in the
deconfined phase, where $\langle L(\vec{x})\rangle \neq 0$. For
QCD, $N =3$ and hence it has three degenerate $Z(3)$ vacua 
resulting from the spontaneous breaking of $Z(3)$ symmetry at $T>T_{c}$. This 
leads to the formation of interfaces between regions of different $Z(N)$ vacua.
These vacua are characterized by,
%
$l(\vec{x}) = 1, e^{2\pi i/3}, e^{4\pi i/3}.$
%
%
\begin{figure}
    \includegraphics[width=0.4\textwidth]{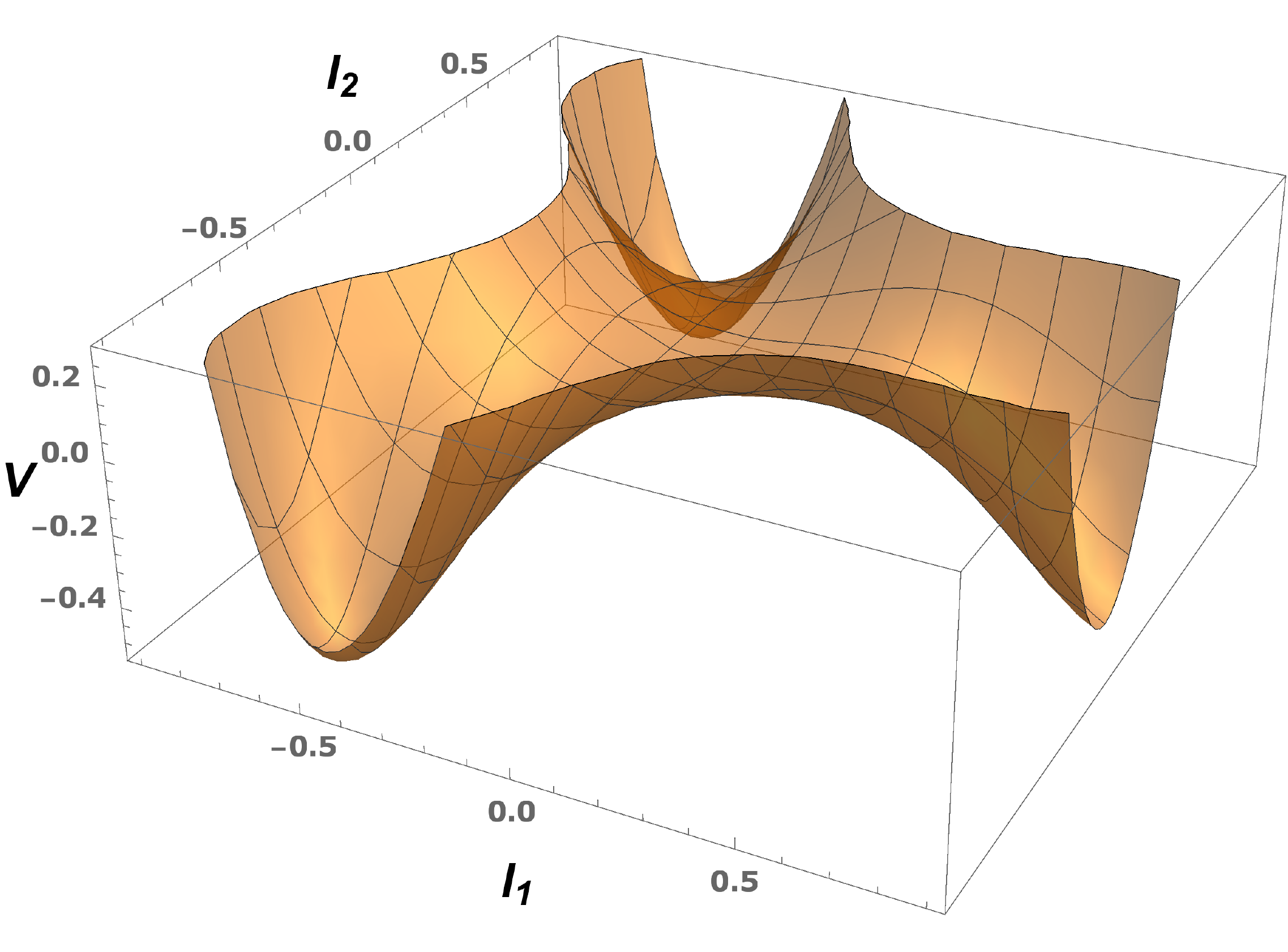}
    \caption{Effective potential of Polyakov loop for $T > T_{c}$. The plot is
    at $T=200$ MeV.}
  \label{fig:pispot}
\end{figure}

The effective potential for Polyakov loop which captures the basic features of
$Z(3)$ structures is given by \cite{Pisarski:2000eq}
\begin{equation}
V(l) = \biggl(-\frac{b_2}{2}|l|^{2} - \frac{b_3}{6}\Bigl(l^{3} 
+ (l^{*})^{3}\Bigr) + \frac{1}{4}(|l|^{2})^{2}\biggr)b_4T^{4},
\label{eq:pispot}
\end{equation}
where $b_2 = \left( 1-1.11/x \right) \left(1+0.265/x\right)^{2}\left(1+0.300/x
\right)^{3} - 0.478$ (with $x=T/T_{c}$ with $T_{c}\sim182$ MeV) ,$b_3 = 2.0$
and $b_4 = 0.6061\times47.5/16$. The value of $T_{c}$ and the
coefficients are fixed by fitting the energy and pressure to the lattice
results \cite{Scavenius:2002ru}. An overall factor of $(47.5)/16$ in $b_{4}$
is used to compensate for the change in the energy density by re-scaling the
number of degrees of freedom for the three flavor case.
In $T\longrightarrow\infty$ limit,
$l\left(x\right)\longrightarrow y = b_{3}/2+\frac{1}{2}\times \sqrt{b_{3}^{2} +
  4b_{2}\left(T=\infty\right)}$. As
$\lvert l(x)\rvert = \exp(-\beta \Delta F)$, $l(x)\longrightarrow 1$ in the
limit $T\longrightarrow\infty$. We re-scale the quantities as
\begin{equation*}
l\left(x\right) \rightarrow \frac{l\left(x\right)}{y}, ~~ b_{2} \rightarrow 
\frac{b_{2}}{y^{2}}, ~~ b_{3} \rightarrow \frac{b_{3}}{y}, ~~ b_{4} \rightarrow 
b_{4}y^{4},
\end{equation*}
to get the desired asymptotic behavior of $l(x)$. Writing
$l(x) = |l(x)|e^{i\theta}$, we see that
$(l^{3} + (l^{*})^{3}) = \cos(3\theta)$. This results in three degenerate $Z(3)$
vacua for $T > T_{c}$, as shown in fig \ref{fig:pispot}.
\subsection{Interaction of Quarks with the Domain Wall}
 \label{sec:qrkint}

We now discuss the interaction of the quarks with the domain wall.
Fig \ref{fig:pispot} shows that $l(x)$ varies as we go from one
vacua to other. As $\lvert l(x)\rvert = \exp(-\beta \Delta F)$
($\Delta F$ being the change in the free energy of a quark); the change in
the free energy of the quark as it traverses the domain wall is given by
\begin{equation}
  \Delta F = -T\ln \left(\lvert l(x)\rvert\right)
  \label{eq:fren}
\end{equation}  

\begin{figure}
  \includegraphics[width=0.4\textwidth]{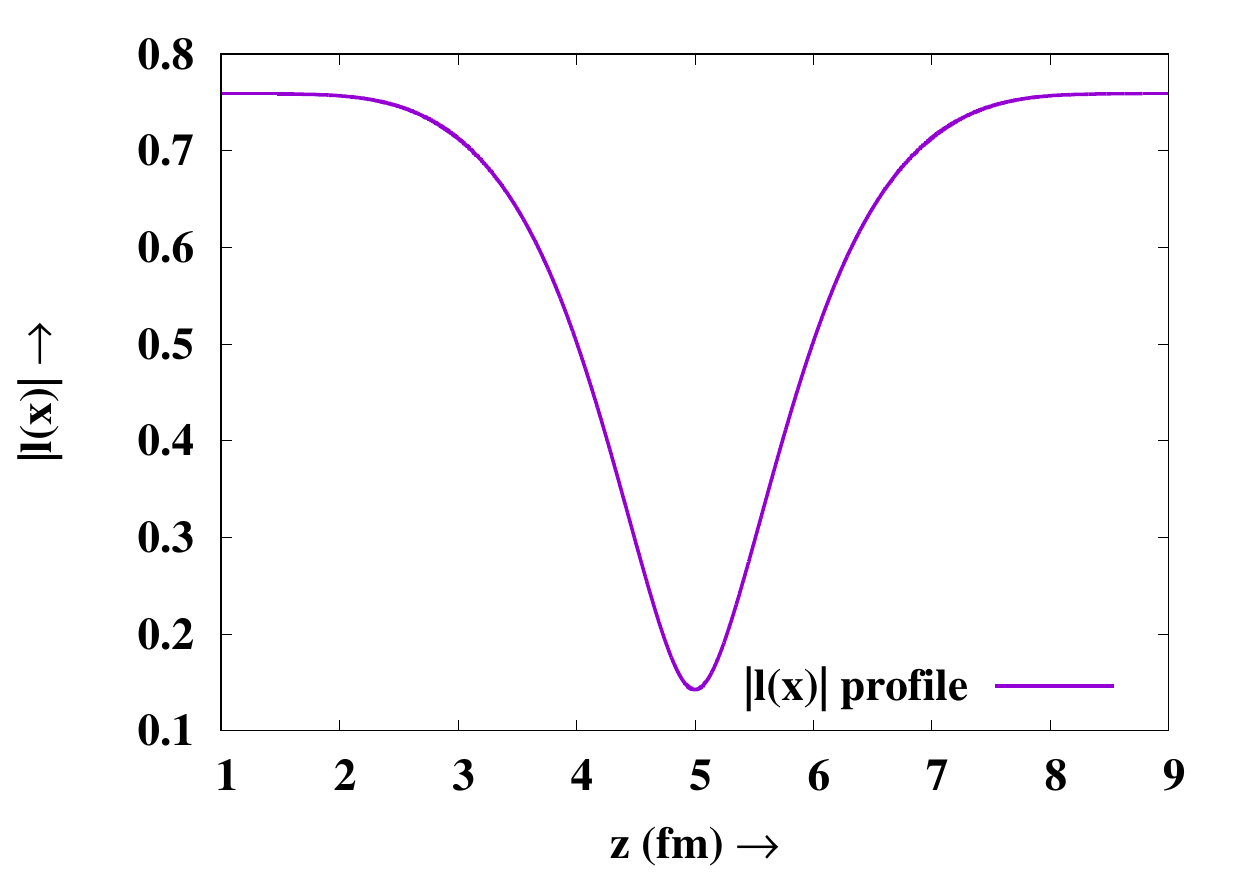}
  \includegraphics[width=0.4\textwidth]{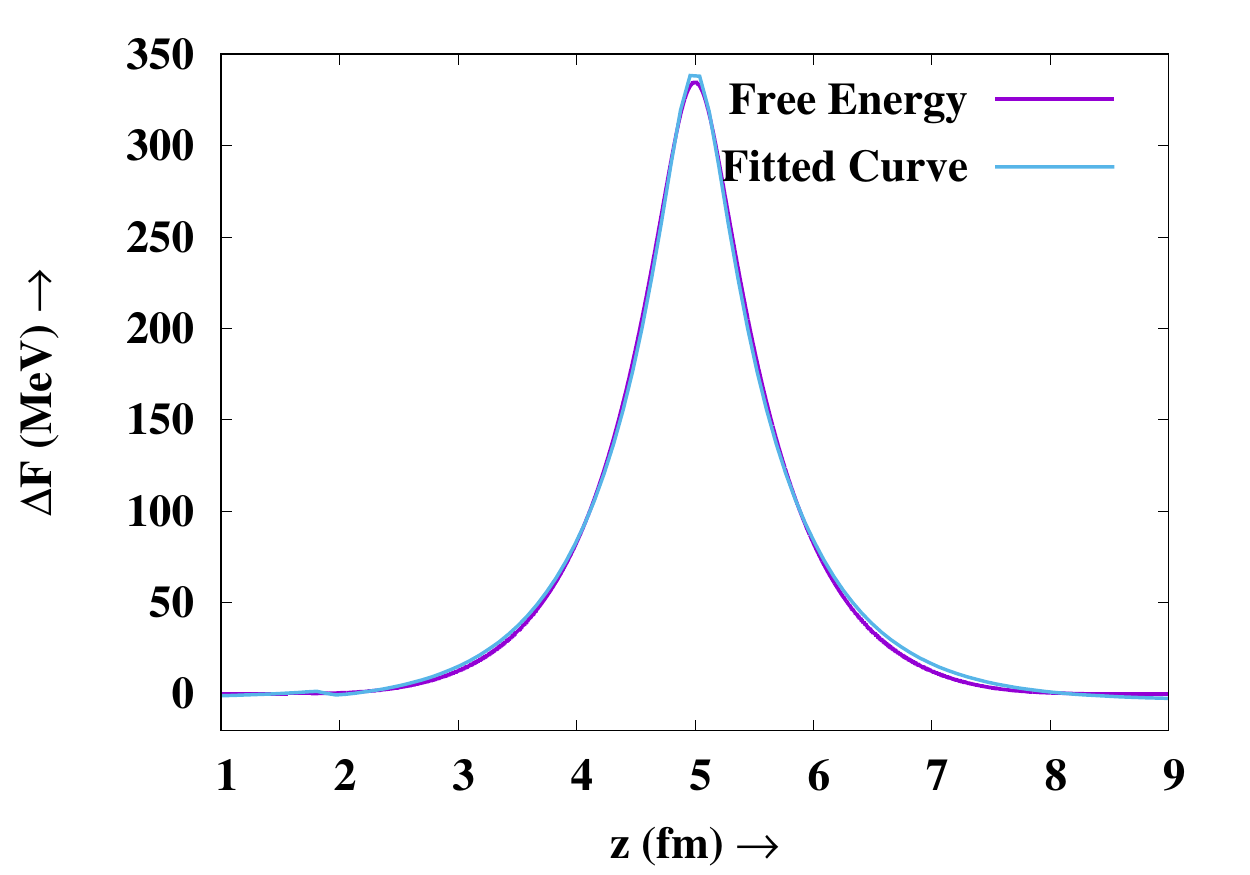}
  \caption{\textit{Top}:- The $l(x)$ profile at $T=200$ MeV.
    \textit{Bottom}:- The corresponding free energy change and the fitted curve
  (see text for details).}
  \label{fig:prfl}
\end{figure}

A moving quark thus experiences an effective potential as it crosses the wall.
To estimate $\Delta F$ we need the domain wall profile $l(x)$. For obtaining
the $l(x)$ profile we use the energy minimization technique. We write,
$l(x) = l_{1}(x)+\dot{\iota} l_{2}(x)$, then we express eq. (\ref{eq:pispot}) in
terms of $l_{1}(x)$ and $l_{2}(x)$. We then start the linear interpolation with
values of $l_{1},l_{2}$  between $\theta = 0$ and $\theta = 2\pi/3$ and
minimize the total energy $\left((\nabla l)^{2} + V(l) \right)$ of the system.
The numerical technique used to minimize energy is over
relaxation technique. This technique requires that the field be fluctuated
at a lattice site and then the change is observed in the energy density due to
the fluctuation. With three such fluctuations a parabola is fitted. The minimum
of the parabola provides the minimum energy configuration. The actual change
in the field is taken as a fraction of this minimum value. We take the the
fraction to be 0.05 times the field value obtained. The energy is then
calculated with the new field values and the process is repeated till the
energy stops changing effectively. The resultant $l(x)$ profile and the
corresponding variation in the free energy are shown in fig \ref{fig:prfl}.
The free energy is fitted with a curve. The form of the function and the
parameters are presented in Table \ref{tab:param}.
For more details on the
process of obtaining $l(x)$ profile, we refer to \cite{Layek:2005fn}.
\begin{table}[!htp]
  \caption{Table for the parameters used for fitting the free energy variation
    with the function
  $-200\ln\lbrace\sqrt{[a\tanh(bx+c)+d]^{2}+[e\tanh(fx+g)+h)]^{2}}\rbrace+p$}
 \label{tab:param}
\begin{center}
\begin{tabular}{lll}
  \hline\noalign{\smallskip}
Parameter & \hspace*{0.2cm}Value \\
\noalign{\smallskip}\hline\noalign{\smallskip}
$\quad \quad a$ & $1.0036$ \\
$\quad \quad b$ & $-0.739001$ \\
$\quad \quad c$ & $3.68957$ \\
$\quad \quad d$ & $0.00450475$ \\
$\quad \quad e$ & $-0.0898081$ \\
$\quad \quad f$ & $-14.4524$ \\
$\quad \quad g$ & $26.833$ \\
$\quad \quad h$ & $0.0916998$ \\
$\quad \quad p$ & $-0.381349$ \\
\noalign{\smallskip}\hline
\end{tabular}
\end{center}
\end{table}

This free energy change is like an effective potential ($V$) experienced by
the quark as it traverses the domain wall. We approximate the $l(x)$ profile
with a box of width $\sim 1$ fm and height $\sim 0.6$. This translates to an
effective potential of width $a=1$ fm and height $\sim 305$ MeV. We use this
potential to estimate the transmission coefficient. The analytic expression for
the transmission coefficient for a free quark incident on a box potential is
given by \cite{Mckellar:1987ru}:
\begin{subequations}
  \label{eq:trans}
  \begin{align}
  T &= \frac{1}{\cos^{2}\left(Ka\right) +
    \frac{1}{4}\Big\lvert\frac{\Lambda}{\lambda}+\frac{\lambda}{\Lambda}
    \Big\rvert^{2}\lvert\sin(Ka)\rvert^{2}}~; \\
  \mathrm{where}\quad \lambda &= \frac{k}{E+m}~;\quad\quad
  ~~\Lambda = \frac{K}{(E-V)+m};\\
  \mathrm{with}\quad k &= (E^{2}-m^{2})^{1/2}~;\quad K=\left((E-V)^{2}-m^{2}\right)^{1/2}
  \end{align}
\end{subequations}
The transmission coefficients for u,d and s
quarks are listed in Table \ref{tab:trans}. From Table \ref{tab:trans}, we
conclude that the transmission coefficients of u/d quarks are higher than the
transmission coefficient of the s-quarks.

The recent lattice results with quarks
\cite{Bhattacharya:2014ara,Steinbrecher:2018phh} have reported slightly
different values of $T_{c}$ than the value used above. It is then advisable to
see how our results obtained above fare in the light of these results. We note
that the
$l(x)$ profile depends on the shape of the potential given in \ref{eq:pispot}.
Since the potential depends on the ratio $T/T_{c}$, it is this value which
will be important, not the exact value of $T_{c}$, in determining the shape of
the potential. For our case we are focusing on $T=200$ MeV, which translates
to $T/T_{c}\sim 1.1$ for $T_c \sim 182$ MeV quoted in ref.
\cite{Scavenius:2002ru}. The shape of the profile will remain same for the
corresponding $T$ which keeps the above value fixed. For $T_{c}\sim 155$ MeV,
which is quoted in ref. \cite{Bhattacharya:2014ara,Steinbrecher:2018phh}, the
corresponding $T\sim 170$ MeV, which is close to $T\sim 200$ MeV used by us.
Thus we hope to capture the correct qualitative features of the $Z(3)$
symmetry in our analysis.

Also replacing a box function with a
smooth profile would change the transmission coefficients. It was shown in
\cite{Atreya:2011wn} that for a step potential, the coefficients are larger
than those for a smooth profile with the width $\sim 1$ fm. This
essentially means that u and d quarks will essentially
pass through and not get reflected while strange quarks will get
reflected but with a lower probability, for a more realistic profile of
the scattering potential. So qualitative features of the
results will not change, as in that the concentration in the collapsing
region would be solely because of strange quarks.
\begin{table}
\begin{center}
\caption{Transmission for different quark flavors for incoming E $=500$ MeV.}
\label{tab:trans}
\begin{tabular}{lll}
  \hline\noalign{\smallskip}
$q$ flavor & $\quad \quad T$\\
\noalign{\smallskip}\hline\noalign{\smallskip}
$u/d$ & $0.999329$ \\
$s$ & $0.855293$ \\
\noalign{\smallskip}\hline
\end{tabular}
\end{center}
\end{table}

\section{Charge Inhomogeneity due to collapsing $Z(3)$ domains}
\label{sec:charge}

In this section we discuss how a net charge is concentrated across the
domain wall as it collapses. We start with a probable scenario of $Z(3)$ domain
formation in early universe and their survivability till the QCD phase
transition. We then estimate the density of quarks trapped inside the
collapsing $Z(3)$ domains. Since quarks have electric charge too, their
concentration leads to a electric charge concentration across
the domain walls. We look at the response of the leptons to this baryon excess
and find that leptons do not cancel this net electric charge concentration.

\subsection{Formation of $Z(3)$ domains in early universe}
\label{sec:z3form}

The standard picture of defect formation in cosmology relies on the Kibble
mechanism \cite{Kibble:1976sj}. The defects are formed during the transition
to the symmetry broken phase as the universe cools during its evolution.
Since $Z(3)$ symmetry is broken in the high temperature phase, and is 
restored as the universe cools while expanding, the question then arises as to
how these defects were formed. This question was first addressed in ref.
\cite{Layek:2005zu}.

The pre inflationary universe was in the deconfined state as $T>>T_{c}$.
During inflation the universe cools exponentially and $Z(3)$ interfaces
disappear as $T\rightarrow 0$. When temperature eventually becomes higher than
$T_{c}$, during reheating, $Z(3)$ symmetry breaks spontaneously, and $Z(3)$ 
defects form via the Kibble mechanism. However
quarks lead to an explicit breaking of $Z(3)$ symmetry. Two of the vacua, with 
$l(x)=z,~z^{2}$, become meta-stable leading to a pressure difference between 
the true vacuum and the meta-stable vacua \cite{Dixit:1991et}. This leads to a
preferential shrinking of the meta-stable vacua.
The explicitly breaking of $Z(3)$ symmetry might get amplified in presence of
magnetic field as discussed in section (\ref{sec:intro}).

As the collapse of these closed regions can be very fast
\cite{Gupta:2010pp,Gupta:2011ag}, they are unlikely to survive until late
times, say until the QCD scale. However the dynamics of these $Z(3)$ walls might
be friction dominated due to the non-trivial scattering of quarks from the
domain wall. For large friction, they might survive till the late times. In
certain low energy inflationary models with low reheating temperature
\cite{Knox:1992iy,Copeland:2001qw,vanTent:2004rc}, even a small friction in 
the domain walls collapse will allow the walls to survive until QCD transition.
If these domains survive till late times (which cannot be below the QCD
transition epoch), then the magnetic fields will be generated near the QCD
transition epoch.

\subsection{Number Density Evolution in the Collapsing Region}
\label{sec:num}

We now estimate the number density concentration within the collapsing region.
We are interested in demonstrating a new method of generating magnetic fields
from collapsing $Z(3)$ domain walls. To make an approximate estimate of the
magnetic field generated in this method, we make some simplifying assumptions.
The first one being
that the collapse happens faster than the Hubble expansion rate, thus the universe
is at a constant temperature. This means that the domain wall configuration can
be taken as constant, since $l(x)$ depends on $T$ which is constant. This also
means that one can neglect the expansion of the universe in the present context.
Another assumption is that the particles thermalize instantaneously as they
reflect from the wall. Finally we ignore the baryon diffusion in the plasma. This essentially
translates to assumption that the baryons instantaneously homogenize after the
collision with the wall.

Let there be $N$ domains with radius $R(t)$. Then the total volume within the
domain is $V_{d} = N(4\pi/3)R(t)^{3}$. If $V$ is the Hubble volume, then
$V-V_{d}\equiv V_{o}$ is the volume outside the region enclosed by domains.
Let the number density of particles outside and inside the closed domains be
$n_{o}$ and $n_{i}$. Then the rate of change of concentration is given by
\begin{figure}[!b]
  \includegraphics[width=0.4\textwidth]{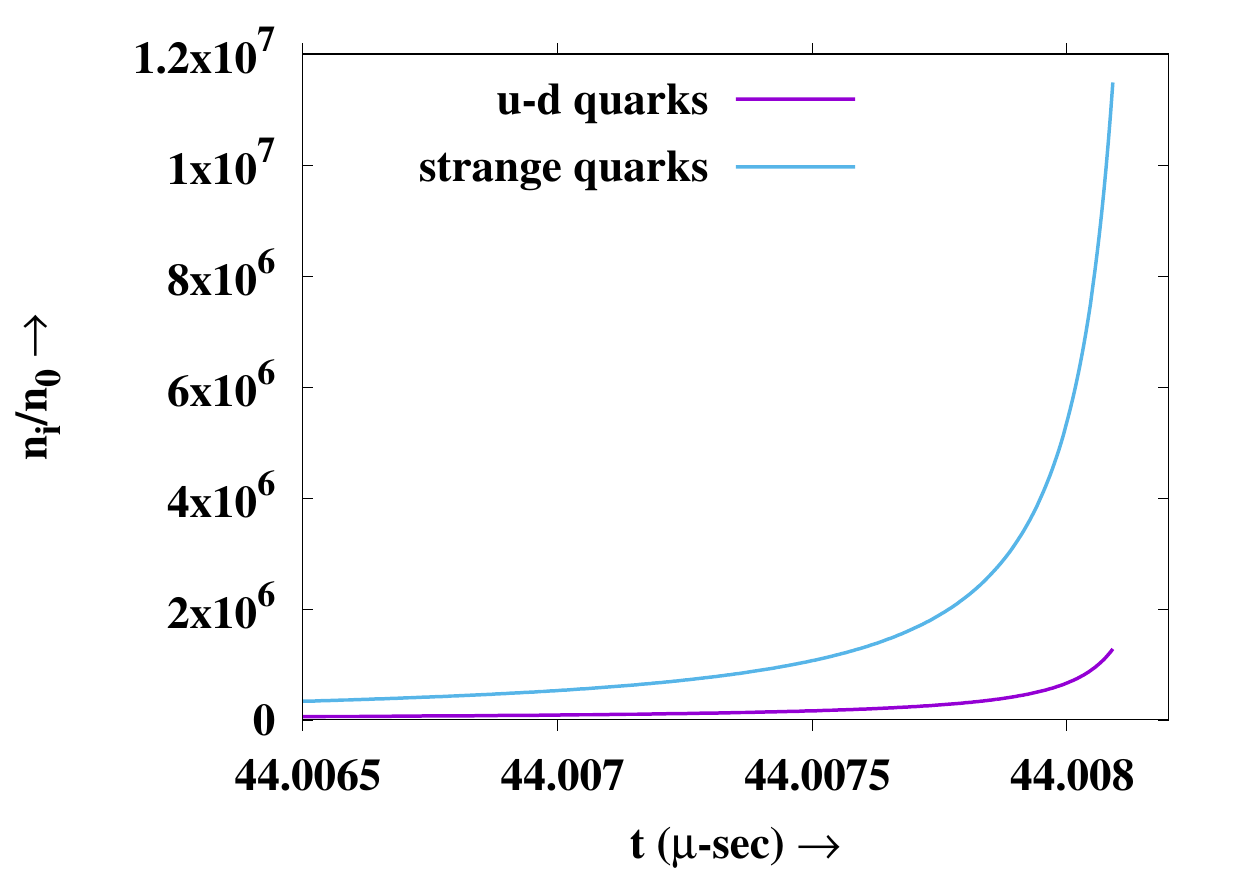}
  \caption{Build up of the quark number density inside the domain wall at
    $T=200$ MeV. The collapse velocity is chosen to be $0.5/\sqrt{3}$.}
  \label{fig:ndens}
\end{figure}
\begin{equation}
\dot{n_{i}} = \Bigl(-\frac{2}{3}v_{w}T_{w}n_{i} + \frac{v_{o}^{rel}n_{o}T_{-}- 
v_{i}^{rel}n_{i}T_{+}}{6}\Bigr)\frac{S}{V_{d}} - n_{i}\frac{\dot{V_{d}}}{V_{d}} \\
\label{eq:numin}
\end{equation}
\begin{equation}
\dot{n_{0}} = \Bigl(\frac{2}{3}v_{w}T_{w}n_{i} - \frac{v_{o}^{rel}n_{o}T_{-}- 
v_{i}^{rel}n_{i}T_{+}}{6}\Bigr)\frac{S}{V_{d}} + n_{o}\frac{\dot{V_{d}}}{V_{o}}, \\
\label{eq:numout}
\end{equation}
where $v_{w}$ is the wall velocity, $S$ is the surface area of the enclosed
region. The particle motion can be studied with components parallel to the
wall $(v_{w})$ and perpendicular to the wall. $v_{o,i}^{rel}$ is the relative
velocity of the particle (outside or inside), perpendicular w.r.t. the wall.
Each particle has $6$ degrees of freedom with $4/6$ in the parallel direction
and $1/6$ moving towards the wall, perpendicular to it. These are the ones that  contribute to the
change in the number density. The remaining $1/6$ are moving away from the wall and therefore do not contribute. The change in the domain volume $\dot{V}_{i}$ is estimated by looking at $R(t)$
\begin{equation}\label{eq:radius}
R(t) = \frac{r_{H}}{N^{1/3}} - v_{w}(t-t_{0}),
\end{equation}
where $r_{H}$ is the horizon size of the universe at $t=t_{0}\simeq 30\left(
\frac{150}{T(MeV)}\right)^{2}$. $T_{w}$ are the reflection coefficient for the
particles moving parallel to wall and $T_{-}(T_{+})$ are the transmission
coefficients calculated for the quarks that are moving from outside (inside)
of the wall towards the inside (outside).

We solve eq. (\ref{eq:numin}), (\ref{eq:numout}) and (\ref{eq:radius})
simultaneously to get the increase in the charge concentration across the
domain wall. We consider the wall velocity to be $0.5/\sqrt{3}$ and the number
of domains within the horizon, $N=10$. The evolution of the number density
normalized to the background number density 
is plotted in Fig. (\ref{fig:ndens}). One can clearly see the rise in the
number density inside the collapsing region. The difference between the
concentration of different quark species is also clearly visible. The entire
concentration inside the domain can be attributed to the strange quarks that
outnumber the other quark species by an order of magnitude. Since the $l(x)$
profile is color neutral, the effective potential as seen by quarks is also
color neutral and thus the anti-quarks have the same concentration.

To estimate the density of quarks trapped in the collapsing domain, we look at
the density profile left behind the collapsing domain. Let $\rho(r)$ be the
particle density left behind the domain wall. Then total number of particles
in a shell of thickness $dr$ at a distance $r$ from the center of domain wall
is given by $dN = 4\pi r^{2}\rho(r)dr$. This gives
\begin{equation}
  \rho(r) = -\frac{\dot{N}}{4\pi r^{2}v_{w}}
  \label{eq:rhoprfl}
\end{equation}
\begin{figure}[!b]
  \includegraphics[width=0.4\textwidth]{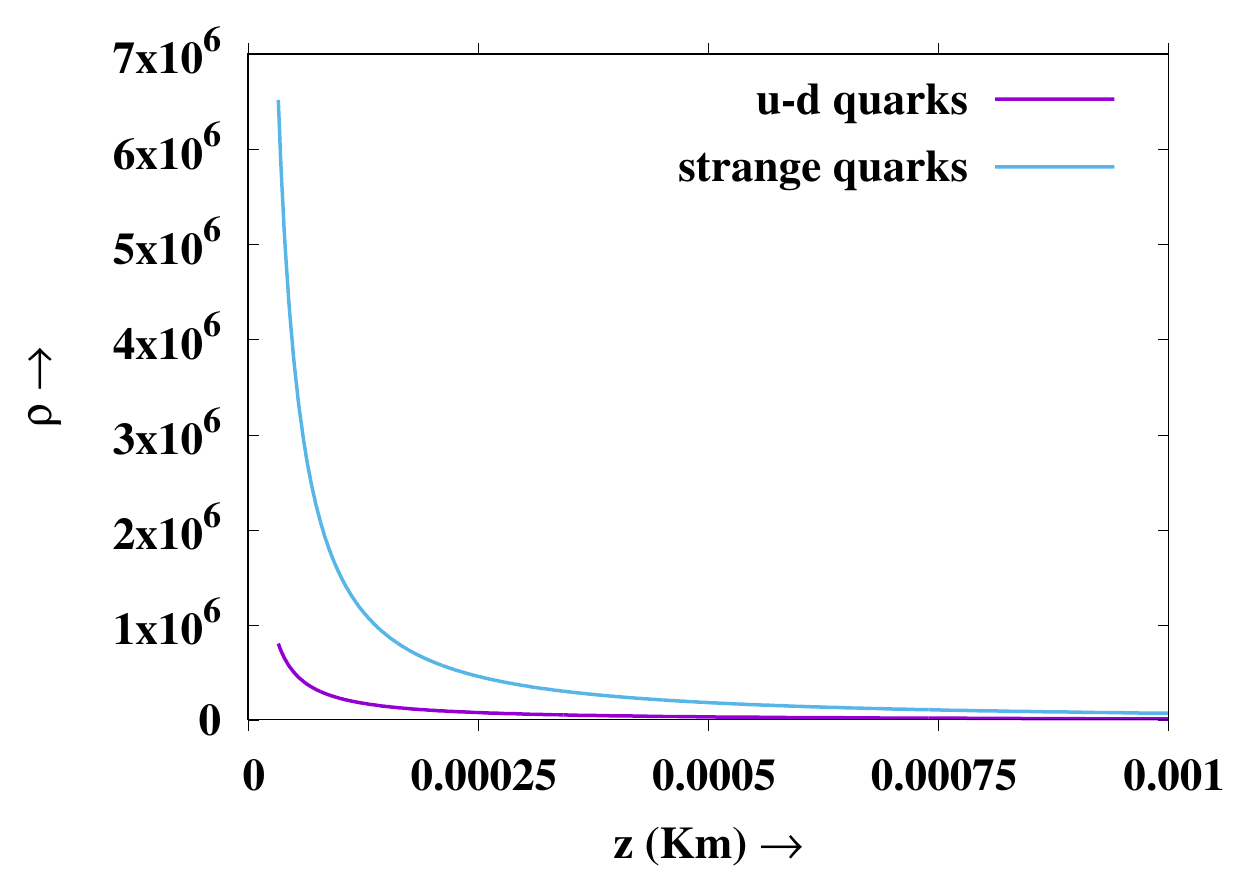}
  \caption{The density profile of the particles left behind by the domain wall
  at $T=200$ MeV. The collapse velocity is chosen to be $0.5/\sqrt{3}$.}
  \label{fig:rho}
\end{figure}
The quark profile as the function of the radius of the collapsing region is
shown in the Fig (\ref{fig:rho}). We  notice that almost the entire quark
concentration can be solely owed to the strange quarks. The same profile will
be obtained for the anti-quarks owing to the color neutral nature of the
effective potential seen by quarks anti-quarks. However there is a small
baryon asymmetry $(\sim 10^{-10})$ in the universe near QCD transition. Thus
even though the density profile is same for the quarks and anti-quarks, there
is a difference in their net numbers. We now make an order of magnitude
estimates of quark over-densities. From Fig (\ref{fig:rho}) we see that the
density within a radius of $1$ m is roughly $10^{6}$ times the background
density.
At $T =200$ MeV, the corresponding length scale is $1$ fm, thus we take the
background density to be $\sim 1$fm$^{-3}$ which is $10^{45}$ m$^{-3}$. Thus we
get $\sim 10^{51}$ quarks within a radius of $1$ m which translates to an
net excess baryon number of $10^{41}$.

The transmission coefficients from eq. (\ref{eq:trans}a) appear directly in
the eq. (\ref{eq:numin}) and (\ref{eq:numout}). Thus naively one can say that
the rate of concentration of quarks will be slower for the realistic scenario
as the transmission is larger there. This would mean that the charge
pile up will take some more time to reach the values quoted in the fig.
(\ref{fig:ndens}). This would translate to the size of domain wall thus
effecting the density profile in fig (\ref{fig:rho}). The numbers quoted
there will be attained a bit later than the present case. However if the
collapse is faster than what we consider it to be (which is likely as argued
in sec \ref{sec:trblnc}), these high densities can be attained quite easily as
$v^{rel}$ and $v_w$ in eq. (\ref{eq:numin}) will compensate for the reduction
in transmission coefficients. In either case the concentration quoted by us
can be attained a bit sooner or later depending on how the realistic
situation unfolds.

\subsection{Dynamical Charge build up across the domain wall}
\label{sec:chrg}

As we discussed, quarks/anti-quarks undergo non-trivial reflection and
transmission from $Z(3)$ domains.
Thus there is a density contrast across the domain wall. We also argued that
anti-quarks do not cancel this baryon excess due to tiny baryon asymmetry.
Since quarks also carry electric charge, a density contrast across the
interface also leads to a charge asymmetry due to the difference in
charges of the u,d and s quarks. While the net baryon over-density is given
by, $n_B = (n_u + n_d + n_s)/3$, the net charge over-density is given by
$\rho_c = e(2 n_u - n_d - n_s)/3$. Substituting, the over-densities of the
respective quarks inside the $Z(3)$ domain walls, we find that
$\rho_c \approx - 10^6 e $.

Due to this charge concentration, a potential is generated across the domain
wall. We can use the Debye model to estimate the strength of the potential. 
We reiterate that since the quark hadron transition has not taken place as
yet, the only charged particles in plasma are the quarks and leptons. 

The strong interaction screening for QGP is effective at the length scale
of $\sim 1$ fm, while it is different for the EM interactions. To estimate the
EM Debye length we make the approximation that near the domain wall the
electrons are moving in the background of quarks and muons which are much
heavier than the electrons. In this approximation the situation is akin to that of
an electron-ion plasma and one can use the following expression to estimate
the Debye length
\begin{equation}
  \lambda_{D} = \Biggl[\frac{1}{2\pi e^{2}\left(\frac{n_{e}}{T_{e}} +
      \sum_{i}\frac{q_{i}^{2}n_{i}}{T_{i}}\right)}\Biggr]^{1/2},
  \label{eq:debye}
\end{equation}
$n_{e}$ and $T_{e}$ are the temperatures of the electron and $q_{i},~n_{i},~T_{i}$ 
are the charge, number density and temperature of the ion species $i$. For us
$T_{e} = T_{i}$ and $q_{i} = 2/3,~ -1/3,~ -1/3$ and $-1$ for $u,~d,~s$ quarks and
muon respectively. Since all particles are relativistic one can assume that
$n_{e}\sim n_{i}$.  This reduces the eq. (\ref{eq:debye}) to
\begin{equation}
\lambda_{D} = \left(\frac{9T}{16 \pi e^{2}n_{e}}\right)^{1/2},
  \label{eq:debye2}
\end{equation}
Since the number density of fermions at temperature $T>>m$, the mass of
electron, is $n_{i} \sim \frac{3 \zeta(3)}{4 \pi^2} T^3$ is (where $\zeta(3)$
is the Riemann zeta function). This translates to $\lambda_{D} \sim 10^{-9}$ m
for $T \sim 200$ MeV. 

\begin{figure}
    \includegraphics[width=0.4\textwidth]{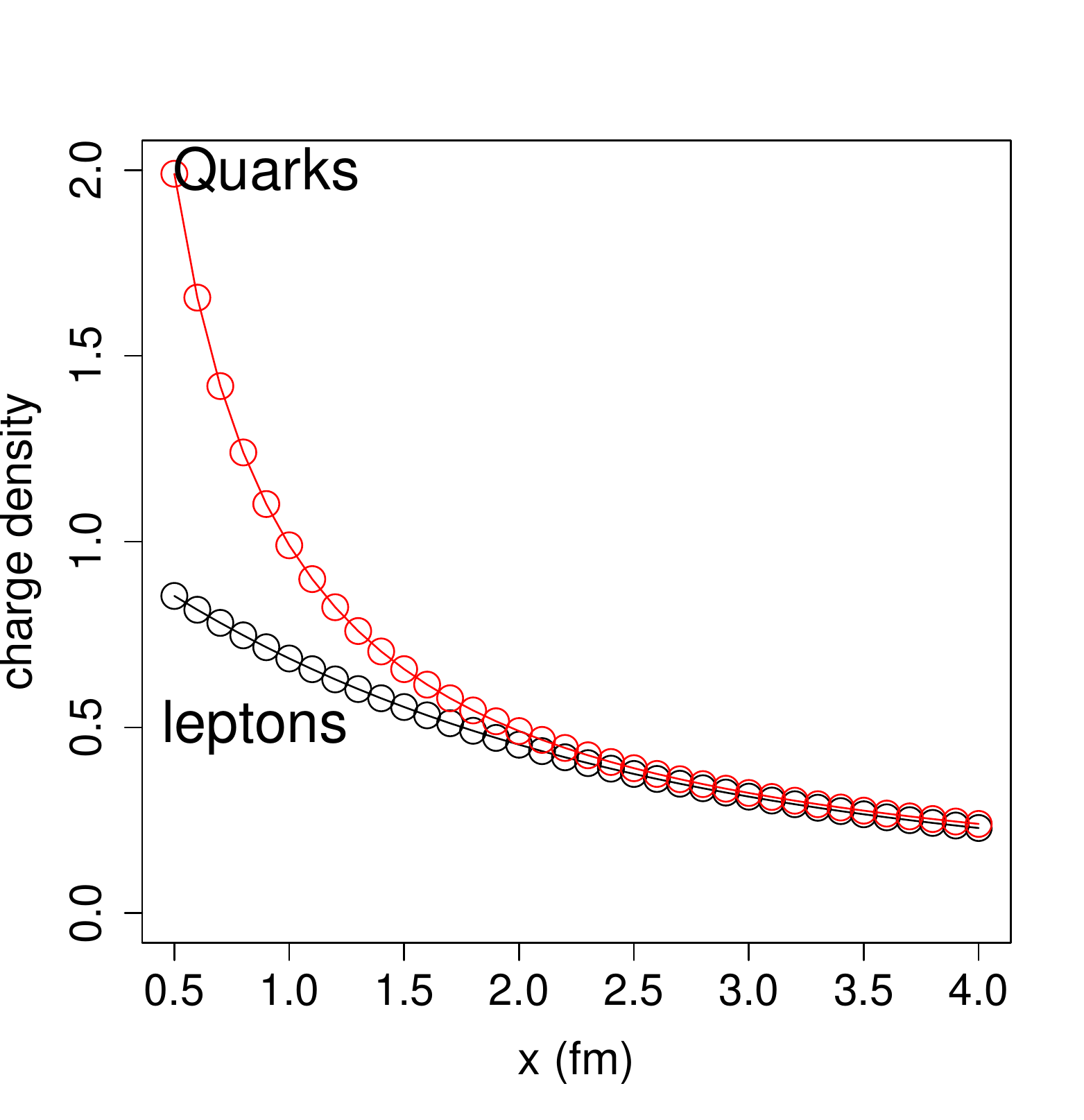}
    \caption{The quark density profile and the lepton density profile
      (scaled by $10^{6}$) near the domain wall.}
  \label{fig:charge}
\end{figure}

The charge density of the quarks on the domain wall has been obtained in the
previous section. For this charge density, one can solve the Poisson's equation
($\frac{\partial^2\phi(x)}{\partial x^2} = -4 \pi \times 10^6 e$) numerically and obtain the potential
$\phi(x)$. Here $x$ is the distance from the domain walls. This gives us the charge density profile for the quarks. Since the potential is now known, one can use the standard solution $n_i e^{(-\frac{e\phi}{T})} $to obtain the lepton profile in response to
the charged potential. Here $n_i$ is the number density of leptons defined previously and $\phi$ is the numerical solution obtained previously. The numerical calculation is done using standard codes in R software. We note that the difference between the two profiles is not very large.
However there exists a finite difference. To show that the lepton charges and
the quark charges cannot cancel each other exactly, we plot the charge density
scaled by a factor of $10^6$ for both the cases.
Fig.\ref{fig:charge} shows the response of leptons to the charge build up of
quarks near the domain wall. We  see that the lepton charge density
doesn't cancel the quark charge density in the vicinity of the domain wall.

\section{Magnetic field generation}
\label{sec:magfld}

In this section we will take a look at the magnetic field generation in the QGP phase
due to the collapse of the $Z(3)$ domain walls.
We first argue that the collapse of the domain walls lead to turbulence in the plasma and
then estimate the strength of the magnetic field due to the charge build up around 
the collapsing domain walls.

\subsection{Turbulence Generation by Collapsing Domains}
\label{sec:trblnc}

 A closed surface
 collapses due to the surface tension $\sigma$, and the energy difference
 between the enclosed region and the outside, $\epsilon$.
 One can then write the
 Lagrangian for the domain wall as \cite{Adams:1989su},
 \begin{equation}
\label{eq:lag}
 L = -4\pi\sigma R^{2}(1-\dot{R}^{2})^{1/2}+\frac{4\pi}{3}R^{3}\epsilon
\end{equation}

 The equation of motion for a perfectly spherical wall is then
 \begin{equation}
\label{eq:eom}
\ddot{R} = \frac{\epsilon}{\sigma}\left(1-\dot{R}^{2}\right)^{3/2} - 
\frac{2}{R}\left(1-\dot{R}^{2}\right).
\end{equation}
 On integrating the above equation we get the velocity of the domain wall as a function of its radius.
\begin{equation}
\label{eq:eom2}
\left(1-\dot{R}^{2}\right)^{1/2} = \frac{R^{2}}{\epsilon R^{3}/3\sigma +
  \mathrm{constant}},
\end{equation}

As the domain wall collapses $R\rightarrow 0$, and $|\dot{R}|\rightarrow 1$.
As a result when a spherical domain collapses, it will produce shocks in the
cosmic plasma. Turbulence is generated in the wake of the shocks. The shocks in
the charge concentration left behind the collapsing region will then generate the magnetic field.

If the dynamics of domain walls is friction dominated, like the scenario above
where the collapse will be slow and for which we have estimated the quark
densities, the shocks will not be produced in the plasma. Even then there
would be turbulence produced near the surface of the domain wall. The momentum equation for the velocity field is given by,
\begin{equation}
  \frac{\partial \vec u}{\partial t} + \vec{u}. \nabla{u} =
  - \frac{\nabla p}{\rho} + \vec{F_{visc}}.
  \label{eq:mom}
\end{equation}
Here $\rho$ is the density of the fluid, $\nabla p $ is the change in
pressure and $\vec {u}$ is the velocity field of the fluid
particles. For simplification, let us assume that the viscous forces
$\vec{F_{visc}}$ are negligible. Due to the concentration of quarks
inside the collapsing walls, there is a jump in the density across the
wall. Therefore, on one side of the domain wall, we have a density
$\rho_{1}$ and on the other side it is $\rho_{2}$.  Since the density
is related to the pressure of the fluid in the two regions, a density
jump indicates a pressure difference $(p_1- p_2)$ across the
wall. Vorticity is generated at an interface or boundary due to a jump
in the tangential acceleration or tangential pressure gradient
\cite{brons:2014}. The exact estimation of the vorticity generated
will require a detailed numerical simulation. In this work we make a
order of magnitude estimates (in the next section) and defer the exact
calculation for future.

In addition, there is a net electric potential produced across the
domain wall since the lepton and quarks do not cancel the electric charge of
each other. This net potential would lead to generation of charged streams of
leptons flowing through each other. There would be two kind of streams for
the positive and the negative charges. The negatively charged leptons would be repelled from
the domain wall in the radial direction (both inward and outward) while the
positive charges would be attracted. The two oppositely charged streams would
flow into each other and lead to the formation of eddies in the cosmic plasma.

\subsection{Estimation of the Magnetic field}
\label{sec:estmt}
We now make an estimate of the magnetic field generated by the collapsing
domain wall. The important thing is to find the charged current in the plasma
which, in turn, requires the knowledge of the velocity of the fluid.
To extract velocity profile one would require detailed numerical simulations
which we presently avoid. Instead, we make an order of estimate of the
vorticity generated. Our method of estimation of the charged current is
similar to the one in ref. \cite{Cheng:1994yr}. Even though the physical
scenario is
quite different, there are some basic similarities: there is an interface
between two regions in the cosmic fluid and the non-trivial reflection of
particles from the interface leads to a density contrast across it.

Far away from the wall the cosmic fluid satisfies the condition 
$\nabla_{\mu}u^{\mu} =0$. Using $u^{\mu} = (\gamma,\gamma \mathbf{v}_{p})$ and
$v^{2}_{p}<<1$, we get $\nabla.\mathbf{v}_{p}= -3H_{QCD}$. Thus
$|\mathbf{v}_{p}| \sim dH_{QCD}$ where $d$ is the distance from the
center of the sphere. In our case, $d$ can be as large as the size of the
domain wall as it starts to collapse. The peculiar velocity generated is therefore 
quite high. The value of $H_{QCD}$ is chosen to be
$t^{-1}_{QCD}\sim \left(10^{-6}\mathrm{s}\right)^{-1}$.

We make an order of magnitude estimate from the magnitude of the current
generated and the peculiar velocity close to the wall.
\begin{equation}
 B \propto (\rho_c \times \rho_{avg}) r_w |\mathbf{v}_{p}|
\end{equation}
Here $\rho_{avg}\sim 10^{-14}$ GeV$^3$ is the average baryon number density at
QCD epoch. $\rho_c \sim 2\times 10^{6}e$ as estimated in the previous section.
To fix the width of the charge layer we consider two cases: $1)$ The effective
charge layer thickness is the Debye length ($\sim 10^{-9}$m) and $2)$ The width
of the charged layer, $r_{w}$ is the diffusion length of quarks in the medium
($\sim \mu$m). This yields $B \sim 10^{15-18}$ G. The estimated magnetic field
thus has a value lower than the equipartition value. However, this is only an
approximate estimate and a better estimate can be obtained by studying the
collapse of the domain wall numerically. 

\section{Conclusion}
\label{sec:discussion}

We have presented a new mechanism of generating a primordial magnetic field in
the early universe from collapsing $Z(3)$ domain walls. The $Z(3)$ domain walls
have different transmission probabilities for the quarks. This gives rise to a
charge asymmetry across the domain wall. This charge asymmetry gives rise to an
electric potential along the sides of the wall. The leptons in the flowing
plasma are affected by this potential. As the walls move through the plasma,
the velocity of the particles changes in the vicinity of the wall. We
discussed how the wall moving in the plasma can lead to vorticity in the
cosmic plasma. The presence of the charge asymmetry and the vorticity near the wall generates the magnetic field.

In this work we describe a novel mechanism to generate magnetic fields near
the QCD phase transition irrespective of the order of phase transition. The
earlier scenarios of magnetic field generation near the QCD phase transition
required a first order Q-H phase transition. However the lattice studies show
that the quark-hadron transition is a cross-over transition for the range of
baryon chemical potential relevant for the early universe. This rules out
essentially all previous models for magnetic field generation at this epoch.
Unlike in the previous cases, where the phase boundary separated two different
phases (QGP and hadrons), we have the same phase (QGP) on both sides of the
domain wall. Thus the vorticity generation and hence the magnetic field
production occurs in the QGP phase itself. Since there are no hadrons in QGP
phase, it is the leptons whose motion is affected by the electric field
generated due to the charge concentration in the vicinity of collapsing domain
wall. We solve the Poisson's equation numerically and show that the lepton
charge is not canceled out by the quark electric charge.

We have not been able to obtain a numerical value for the velocity generated
due to the vorticity and this is a limitation that we would like to work upon.
However we made an order of magnitude estimate of the peculiar velocities
generated by the collapsing domain and used these estimates to
obtain the strength of the magnetic field thus produced. Simulations of $Z(3)$
show that the collapsing $Z(3)$ domains would give rise to shock waves. In
addition, the motion of charged particles across the domain wall could lead to
a two stream instability in the plasma as well. Also there could be other
hydro-dynamical instabilities if the collapse of the domain wall is
non-spherical.
The presence of shock waves and instabilities will definitely contribute to the
generated magnetic field. We have avoided any such complications and argued
that even in the simplest of scenarios a large magnetic field can be generated
in the early universe. Further investigations, incorporating the shock and
instabilities, are needed to give us a better understanding of the nature of
the magnetic field generated from the collapsing domain walls.

\section*{Acknowledgment}
AA would like to thank Rajarshi Ray and Maya P.N. for comments and
useful discussions. AA is financially supported by Scientific Education and
Research Board (SERB), under the National Post-Doctoral Fellowship (NPDF)
grant number PDF/2017/000641.

\end{document}